\documentclass{article}
\usepackage{authblk}
\usepackage{fullpage}
\usepackage{parskip}
\usepackage{array}
\usepackage{graphicx}
\usepackage{hyperref}
\usepackage[font=small,labelfont=bf]{caption}
\usepackage[maxbibnames=8,url=false,doi=false,isbn=false]{biblatex}
\begin{document}
\title{Kunyu: A High-Performing Global Weather Model Beyond Regression Losses}
\author[1]{Zekun Ni}
\affil[1]{Accutar Biotechnology Inc, mike.zekun@gmail.com}
\maketitle
\begin{abstract}
	Over the past year, data-driven global weather forecasting has emerged as a new alternative to traditional numerical weather prediction. This innovative approach yields forecasts of comparable accuracy at a tiny fraction of computational costs. Regrettably, as far as I know, existing models exclusively rely on regression losses, producing forecasts with substantial blurring. Such blurring, although compromises practicality, enjoys an unfair advantage on evaluation metrics. In this paper, I present Kunyu, a global data-driven weather forecasting model which delivers accurate predictions across a comprehensive array of atmospheric variables at $0.35^\circ$ resolution. With both regression and adversarial losses integrated in its training framework, Kunyu generates forecasts with enhanced clarity and realism. Its performance outpaces even ECMWF HRES in some aspects such as the estimation of anomaly extremes, while remaining competitive with ECMWF HRES on evaluation metrics such as RMSE and ACC. Kunyu is an important step forward in closing the utility gap between numerical and data-driven weather prediction.
\end{abstract}
\section{Introduction}
Weather forecasts play a crucial role in helping human beings plan their activities and avoid disasters. Conventionally, these forecasts rely primarily on global numerical weather prediction (NWP) models such as ECMWF HRES. Decades of evolution has cemented NWP as a reliable and precise method for weather forecasting. However, they are very computationally expensive and often necessitates hours for completion on supercomputers. Furthermore, systemic errors can be introduced by parametrization schemes \cite{beljaars2018}, issues that can prove difficult to address.

In recent years, with the rise of artificial intellilgence, researchers have increasingly embraced a data-driven approach in weather forecasting, which is now recognized as deep learning weather prediction (DLWP). One of the pioneering contributions in this field was made by Weyn et al.\cite{weyn2019}, although their methodology featured a modest resolution of $2.5^\circ$ and included no more than three variables per grid. Rapid progress has taken place over the past two years. FourCastNet\cite{kurth2023fourcastnet} pushes model resolution to $0.25^\circ$, incorporating nearly two dozen variables per grid. Pangu-Weather\cite{bi2023pangu_weather} is the first to claim superiority over ECMWF HRES based on RMSE evaluation. GraphCast\cite{lam2023graphcast} introduces autoregressive training and surpasses ECMWF HRES on most evaluation metrics using a stricter evaluation strategy. FengWu\cite{chen2023fengwu} and FuXi\cite{chen2023fuxi} extend the length of autoregressive training even further, producing even more favorable evaluation results.

Despite the notable advancements, these models uniformly employ regression losses in the training process irrespective of their choices on architecture. As a result, a significant blurriness is present in their forecasts, evident in their publications. This blurriness progressively worsens with an increasing number of autoregressive steps. In fact, training models on regression losses can be viewed as learning the expected value of a random variable, which is a departure from the principle of existing NWP models. Instead of predicting the expected value, traditional NWP models only produce \textit{one} probability. An analogy exists in NWP prediction where one can compute the expected value by averaging all ensemble members in \textit{ensemble weather forecasting}. However, the outputs of ensemble members offer much more information than the mean value alone, covering topics such as the likelihood of extreme heat or cold or tropical cyclone development. This underscores the inherent limitation of the prevailing DLWP training methodology. Arguably, existing DLWP models should not be benchmarked against deterministic NWP models but rather the mean value derived from ensemble NWP forecasts. As of now, superior performance against ECMWF ENS remains an elusive achivement as demonstrated in FuXi's paper\cite{chen2023fuxi}.

In this paper, I develop Kunyu, the first DLWP model that goes beyond the conventional reliance on regression loss. Characterized by a hybrid structure with spherical convolution and Swin Transformer blocks as well as a training process incorporating adversarial losses, Kunyu more closely resembles ECMWF HRES in both numerical processes and output quality. The model produces fine-grained 6-hourly forecasts across a comprehensive array of atmospheric variables at $0.35^\circ$ resolution for up to 10 days. Evaluations reveal that Kunyu not only aligns with ECMWF in terms of RMSE and ACC, but also outperforms in some aspects, such as the estimation of anomaly extremes and probably also genesis and development of tropical cyclones. At the same time, Kunyu enjoys the same characteristic advantages as other DLWP models, which are blazingly fast inference and minimal resource requirements. It takes less than 6 minutes to finish a 10-day forecast run on a single NVIDIA L4 GPU. Consequently, Kunyu is much closer than existing DLWP models towards a drop-in replacement for NWP models.
\section{Methodology}
\subsection{Overview}
Similar to previous studies, Kunyu is designed to follow the concept of a functional transformation which converts the weather state at time $t$ to the predicted weather state at time $t+\Delta t$, where $\Delta t$ is defined as 6 hours. Subsequent forecasts beyond the initial prediction are carried out through an iterative process, where the predicted weather state at the previous time step is used as the input of the successive time step. Practically, it means predictions for time $t+k\Delta t$ are carried out by applying Kunyu $k$ times to the input weather state at time $t$.

In the representation the weather state, I follow the methodology of FourCastNet, adopting a regular lat-lon grid where variables at different levels become independent channels. Owing to constraints imposed by the architecture of the discriminator model as well as resource limitations, the weather state is modeled on a $512\times 1024$ grid instead of a $0.25^\circ$ ($721 \times 1440$) grid which is the conventional choice in previous research. There are 195 variables participating in the iterative inference process on each grid point.
\subsection{Generator Architecture}
Kunyu follows a U-Net\cite{ronneberger2015} backbone structure, featuring three downsampling and upsampling passes, therefore establishing four resolution levels. On each resolution level however, convolution blocks in U-Net are replaced with alternating spherical convolution and Swin Transformer blocks, interconnected by skip connections.
\paragraph{Spherical convolution block} This block consists of a sequence of operations, including a spherical harmonics transform into spectral space, followed by a ``convolution'' within the spectral domain, and concluding with a spherical harmonics transform back to grid space. The convolution process is based on the sifting convolution proposed in \cite{roddy2021}, while restricting the rank of $g_{lm}$ matrix to 1, a measure aimed at reducing the number of parameters. The introduction of spherical convolution blocks makes Kunyu iteratively transform the input tensor between spectral and grid spaces, which resembles the numerical procedures employed in ECMWF IFS \cite{ecmwf2023}.
\paragraph{Swin Transformer block} This block follows the ordinary Swin Transformer block described in \cite{liu2021}. For higher resolutions, a window size of $8\times16$ is used to accomodate resource constraints, while for lower resolutions a larger window size of $16\times32$ is adopted.
\subsection{Discriminator Architecture}
The discriminator consists only of downsampling passes which progressively reduce the size of input all the way to $2\times4$. On each resolution level the discriminator structure closely mirrors the generator by featuring alternating spherical convolutions and grid space transformations. Recognizing the computational cost of Transformer-based discriminators with marginal practical benefits \cite{dubey2023}, I opted for a similar approach to StyleSwin's\cite{zhang2022}. In my approach, the Swin Transformer blocks in the generator are replaced with modified grid convolution blocks in the discriminator as the transformations carried out within grid domain. Additionally, same as StyleSwin, a wavelet component is incorporated into the discriminator in order to suppress artifacts induced by windowed attention.
\paragraph{Grid convolution block} A ``grid convolution'' actually involves multiple convolutions utilizing the same kernel but different dilations along the longitudinal axis to account for resolution variations between lower and higher latitude zones. To reduce computational overhead, convolutions with larger dilations are selectively computed at higher latitudes. The output is then derived through a weighted sum of these convolutions with learned latitude-dependent weights.
\subsection{Dataset}
The primary source of the training dataset is the ERA5 archive\cite{hersbach2020}. Years from 2000 to 2017 are used in training, while the year 2018 is reserved for evaluation. Unlike conventional practices, atmospheric variables are selected from a 37-level subset of ECMWF model levels instead of the default 37 pressure levels available in ERA5. The variables selected for training in Kunyu are listed in the table below.
\begin{center}
	{\renewcommand{\arraystretch}{1.5}\begin{tabular}{|l|l|}
		\hline
		Model level variables&z,t,r,u,v\\\hline
		Surface variables&sp,skt,src,stl1,swvl1,istl1,sd,rsn,tsn,ci\\\hline
		{\renewcommand{\arraystretch}1\begin{tabular}{@{}l@{}}Surface variables\\(output-only)\end{tabular}}&2t,2d,10u,10v,tcc,hcc,mcc,lcc,tp,sf\\\hline
	\end{tabular}}
\end{center}
It's worth noting that variables listed in the last row are produced at output without being iteratively fed back into the model. Additionally, for variables tp and sf, they are substituted with precipitation data from the GPM dataset\cite{huffman2019final}\cite{huffman2019late} whenever such data is available. Besides atmospheric and land variables, temporal information, including sun longitude, earth-sun distance, hour angle and solar zenith angle, as well as terrain information such as surface altitude, vegetation and soil types, are incorporated into the model input.
\subsection{Training}
\paragraph{Generator} The training objective for Kunyu, the generator, involves a hybrid of regression and adversarial losses as defined in the formulas below:
$$L_G=L_{MSE}+\alpha L_{adv};$$
$$L_{MSE}=\frac1{|X|}\sum_{\{x_{t+k\Delta t}\}_{k=0}^T\in X}\frac1{T|G|}\sum_{k=1}^T\sum_{v\in V}w_v\sum_{g\in G}|g|f\left(\frac{\hat x^{v,g}_{t+k\Delta t}-x^{v,g}_{t+k\Delta t}}{\sigma_{v,k}}\right).$$
Here, $x_t$ represents the weather state at time t, $\hat x_t$ is the predicted weather state, $T$ is the number of autoregressive steps, $G$ denotes the grid, $V$ denotes the variable set, $w_v$ is the weight assigned to the variable $v$, $\sigma_{v,k}$ is the per-variable standard derivation of the difference between weather states at time $t+k\Delta t$ and $t$ (for output-only variables, $\sigma_{v,1}=\sigma_v$ is the standard derivation of that variable, and $\sigma_{v,k}=\infty\,\forall k>1$), $f(x)=200\log(1+x^2/200)$, $\alpha=0.05$; and
$$L_{adv}=-\frac1{|X|}\sum_{\{x_{t+k\Delta t}\}_{k=0}^T\in X}\frac1T\sum_{k=1}^T\log(D(\hat x_{t+(k-1)\Delta t},\hat x_{t+\Delta t})).$$
\paragraph{Discriminator} The discriminator is trained with cross entropy loss and $R_1$ regularization\cite{mescheder2018} with $\gamma=0.1$:
$$L_D=-\frac1{|X|}\sum_{\{x_{t+k\Delta t}\}_{k=0}^T\in X}\frac1T\sum_{k=1}^T(\log(D(x_{t+(k-1)\Delta t},x_{t+k\Delta t}))+\log(1-D(\hat x_{t+(k-1)\Delta t},\hat x_{t+k\Delta t})))+L_{R_1}.$$
Spectral normalization\cite{miyato2018} is further applied to the discriminator, with a slight relaxation to cap the spectral norm at 2, inspired by \cite{lee2022} where the norm remains consistent with the initial norm. This adjustment is necessary to ensure the convergence of adversarial training.
\paragraph{Training schedule} The training process unfolds across several phases.
\begin{itemize}
	\item[(1)] Initially the generator is trained exclusively with regression loss with $T$ set to 1.
	\item[(2a)] An all-regression version named \textbf{Kunyu-Legacy} diverges from Kunyu at this point and undergoes fine-tuning solely with regression loss, with $T$ set to 4.
	\item[(2)] The generator continues training only with regression loss, while the discriminator starts training using samples from the generator. This phase is brief, spanning only a few hundred batches.
	\item[(3)] Adversarial training is subsequently initiated with $T$ staying at 1. For each generator update, the discriminator undergoes 4 updates.
	\item[(4)] Finally, $T$ is switched to 4, and a replay buffer, similar to the one described in FengWu's training process\cite{chen2023fengwu}, is deployed to enhance model stability at longer lead times. The weight of the regression loss is decreased for examples coming from the replay buffer (i.e. those with long lead times) to prevent regression effects from spiraling out of control. This refined model is termed \textbf{Kunyu}.
\end{itemize}
The whole process takes three to four weeks on a single instance equipped with 8 NVIDIA A100 GPUs. Remarkably, the all-regression variant Kunyu-Legacy concludes training in a mere 8 days. The accelerated convergence should be attributed in part to the introduction of spherical convolution blocks, leading to significant savings in computational resources and therefore costs. It can't be ruled out that, owing to resource constraints, the models may not attain full convergence.

The training code is implemented within the PyTorch framework, leveraging the SHTns library\cite{schaeffer2013} for spherical harmonics transforms. Accomodating both the generator and discriminator models within a single 80GB GPU requires significant optimizations, including mixed-precision training, extensive gradient checkpointing and some other tricks which involve modifications to certain PyTorch codes.
\section{Results}
The evaluation strategy follows previous studies, where all models are evaluated on the data from the year 2018. For DLWP models including Kunyu and Kunyu-Legacy, the ERA5 dataset serves as both the initial weather state and the ground truth. On the other hand, ECMWF HRES undergoes evaluation against its own analyses at the corresponding valid times. To reduce the time and computational resources required in the evaluation process, a subset of 142 initial conditions, evenly distributed in time, is selected from the dataset. ECMWF HRES and Pangu-Weather evaluations are only conducted on 24-hour time steps, while other models are evaluated on 6-hour steps. Anomalies are calculated relative to the 30-year ERA5 average spanning from 1990 to 2019.
\subsection{Quantitative Skill}
\begin{figure}[!t]
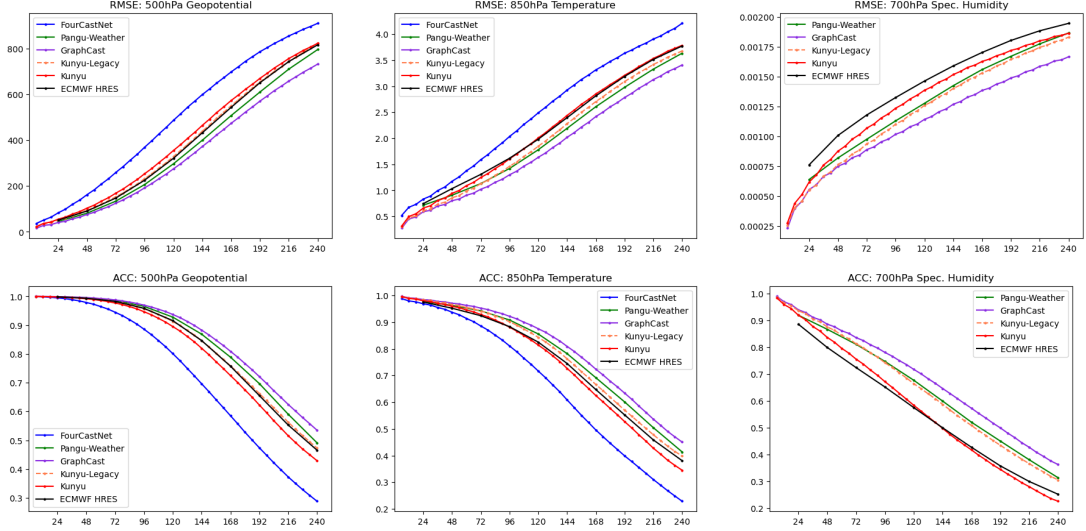

	\centering
	\begin{tabular}{ccc}
		\includegraphics[height=3.5cm]{rmse\_z500.png}&\includegraphics[height=3.5cm]{rmse\_t850.png}&\includegraphics[height=3.5cm]{rmse\_q700.png}\\
		\includegraphics[height=3.5cm]{acc\_z500.png}&\includegraphics[height=3.5cm]{acc\_t850.png}&\includegraphics[height=3.5cm]{acc\_q700.png}
	\end{tabular}
	\caption{RMSE and ACC skills of Kunyu and Kunyu-Legacy across three variables z500 (unit: $\mathrm m^2/\mathrm s^2$), t850 (unit: K) and q700 (unit: kg/kg), in comparison with those of ECMWF HRES and other DLWP models.}
	\label{fig:skill}
\end{figure}
\autoref{fig:skill} illustrates the RMSE and ACC forecast skills for Kunyu and Kunyu-Legacy, alongside ECMWF HRES and other DLWP models, on three pivotal upper-air variables: 500hPa geopotential (z500), 850hPa temperature (t850) and 700hPa specific humidity (q700).

Kunyu-Legacy, the all-regression version of Kunyu, exhibits superior performance against ECMWF HRES across most metrics and closely tracks, albeit slightly trailing, Pangu-Weather. This performance discrepancy can be attributed to the training strategy: Kunyu-Legacy, with its 4-step autoregressive training scheme, averages losses from all four steps. In contrast, Pangu-Weather, with a time step of 24 hours, can be perceived as effectively trained solely on the loss from the fourth step, therefore benefiting more from regression. Generally speaking, Kunyu-Legacy demonstrates competitive modeling capabilities of Kunyu's architecture, definitely surpassing FourCastNet and aligning closely with Pangu-Weather, and potentially GraphCast as well.

Following the introduction of adversarial training however, an obvious albeit modest decline in forecast skill scores can be observed, with Kunyu falling behind Kunyu-Legacy across all metrics. This toll is anticipated and largely unavoidable, but the introduction of regularizations in the discriminator curbs the extent of performance degradation. Compared with the non-regression weather model ECMWF HRES, Kunyu closely matches its performance on RMSE, albeit marginally inferior on ACC. The most pronounced disparities between Kunyu and regression-based models emerge in specific humidity, where the extent of smoothness in regression-based model outputs is most evident. The RMSE and ACC curves of Kunyu on 700hPa specific humidity lag behind other DLWP models but trend most closely with ECMWF HRES.
\subsection{Visualization}
\begin{figure}
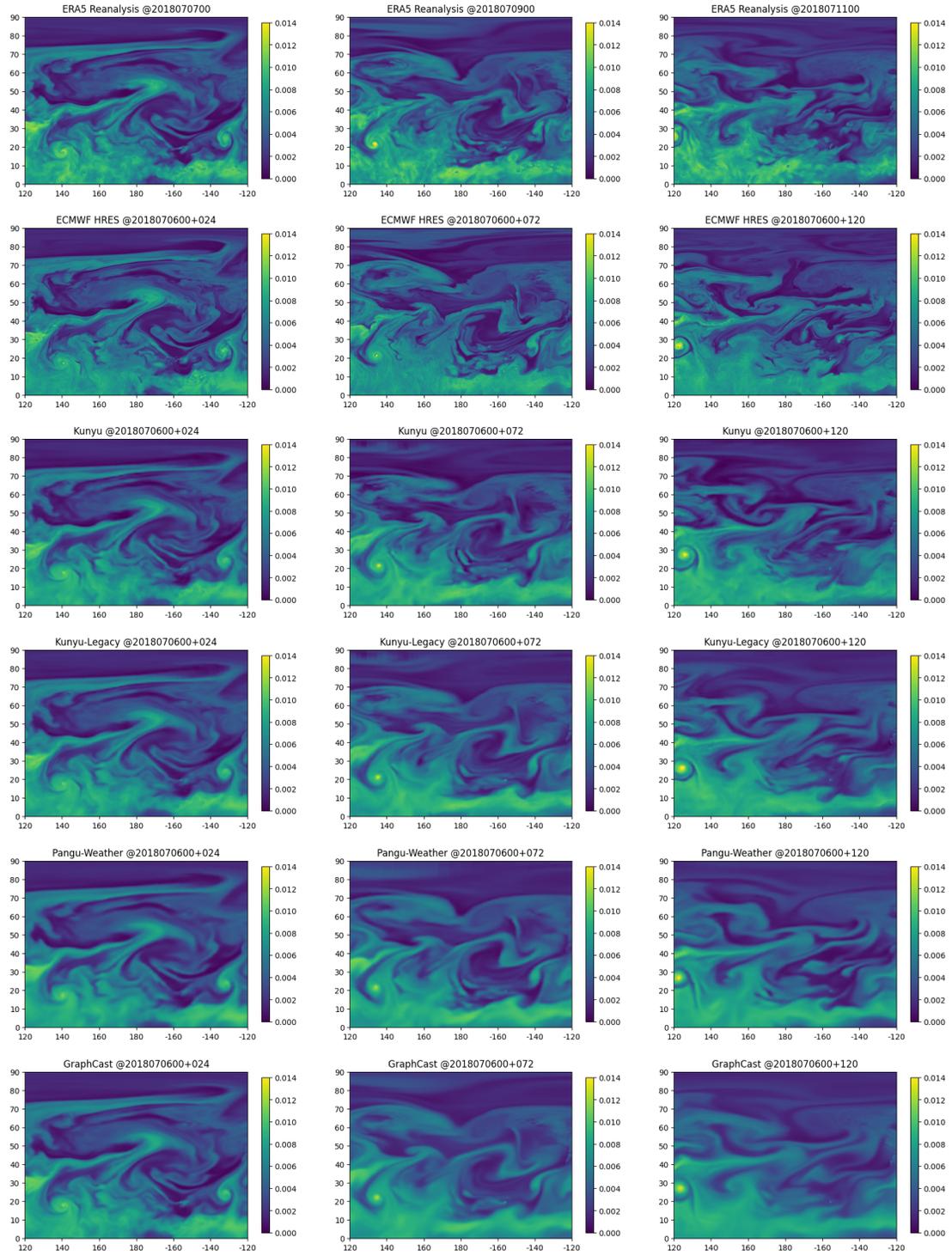

	\centering
	\begin{tabular}{ccc}
		\includegraphics[width=4.6cm]{q700\_era5\_024.png}&\includegraphics[width=4.6cm]{q700\_era5\_072.png}&\includegraphics[width=4.6cm]{q700\_era5\_120.png}\\
		\includegraphics[width=4.6cm]{q700\_hres\_024.png}&\includegraphics[width=4.6cm]{q700\_hres\_072.png}&\includegraphics[width=4.6cm]{q700\_hres\_120.png}\\
		\includegraphics[width=4.6cm]{q700\_kunyu\_024.png}&\includegraphics[width=4.6cm]{q700\_kunyu\_072.png}&\includegraphics[width=4.6cm]{q700\_kunyu\_120.png}\\
		\includegraphics[width=4.6cm]{q700\_kunyu\_legacy\_024.png}&\includegraphics[width=4.6cm]{q700\_kunyu\_legacy\_072.png}&\includegraphics[width=4.6cm]{q700\_kunyu\_legacy\_120.png}\\
		\includegraphics[width=4.6cm]{q700\_pangu\_024.png}&\includegraphics[width=4.6cm]{q700\_pangu\_072.png}&\includegraphics[width=4.6cm]{q700\_pangu\_120.png}\\
		\includegraphics[width=4.6cm]{q700\_graphcast\_024.png}&\includegraphics[width=4.6cm]{q700\_graphcast\_072.png}&\includegraphics[width=4.6cm]{q700\_graphcast\_120.png}
	\end{tabular}
	\caption{Visualization of 24-, 72- and 120-hour forecasts for 700hPa specific humidity by Kunyu, Kunyu-Legacy and other DLWP models, initialized with ERA5 Reanalysis data at 00 UTC on July 6, 2018, in comparison with ECMWF HRES and ground truth data from ERA5. Results shown are focused on the North Pacific area ($120^\circ$E--$120^\circ$W, 0--$90^\circ$N).}
	\label{fig:visual}
\end{figure}
Visual comparison is the most intuitive way to find out the key distinctions between generative models and regression models, with specific humidity variables showcasing the most pronounced effect of smoothness in regression models. In \autoref{fig:visual}, I present plots of 700hPa specific humidity forecasts for various lead times from different models, together with ERA5 reanalysis data as the ground truth for visual assessment.

In general, the level of smoothness increases from top to bottom (with the exception of ECMWF HRES) and from left to right. Predictions from regression models trained with longer time steps such as GraphCast tend to lose most details, while Kunyu's outputs unequivocally emerge as the sharpest among all DLWP models. Differences may be less evident at 24-hour lead time, especially with Kunyu-Legacy, but become progressively more prominent as lead time grows.

Comparisons with ERA5 ground truth reveal that, while Kunyu still exhibits a slight smoothness effect attributed to the presence of regression losses in training, the discriminator keeps this effect in check, particularly at long lead times. The most notable visual differences lie in the tropics, where ERA5 exhibits the texture of scattered convection hotspots, a feature mostly absent in Kunyu's outputs, but it's worth noting that such texture is also largely absent in ECMWF HRES's outputs. Conversely, differences in higher latitudes are less distinguishable.
\subsection{Extremes}
\begin{figure}[!t]
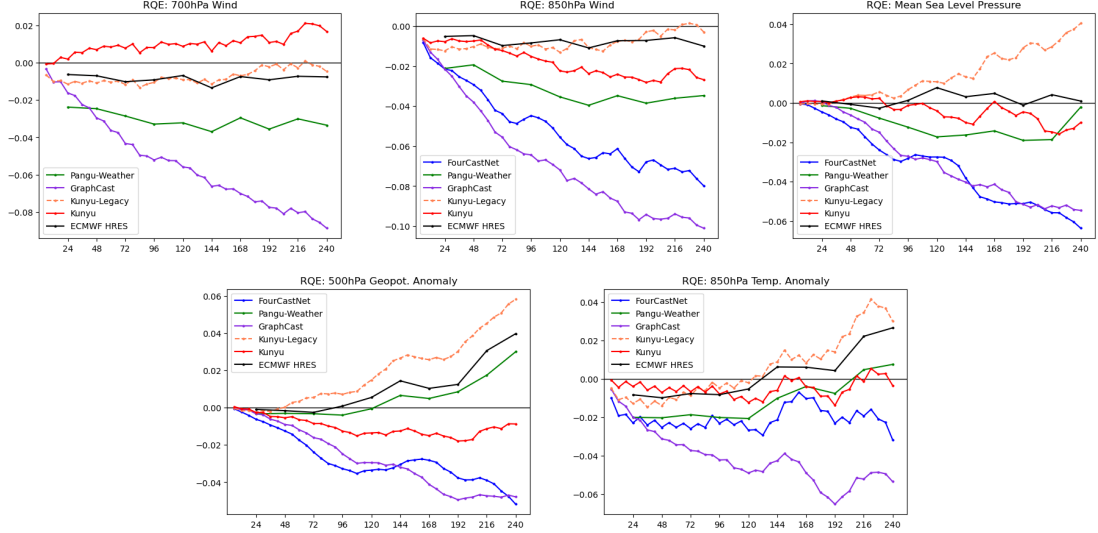

	\centering
	\begin{tabular}{ccc}
		\includegraphics[height=3.5cm]{rqe\_wind700.png}&\includegraphics[height=3.5cm]{rqe\_wind850.png}&\includegraphics[height=3.5cm]{rqe\_mslp.png}
	\end{tabular}
	\begin{tabular}{cc}
		\includegraphics[height=3.5cm]{rqe\_anom\_z500.png}&\includegraphics[height=3.5cm]{rqe\_anom\_t850.png}
	\end{tabular}
	\caption{RQE values of Kunyu and Kunyu-Legacy on 700hPa and 850hPa wind speed, mean sea level pressure and z500 and t850 anomalies, in comparison with those of ECMWF HRES and other DLWP models.}
	\label{fig:rqe}
\end{figure}
I employ the RQE metric proposed in \cite{kurth2023fourcastnet} to assess models' ability in forecasting extreme weather, with a slight modification excluding areas north of $60^\circ$N and south of $60^\circ$S. This modification is made due to small grid areas in those regions, so that the scores are prevented from being skewed too much towards these areas. The RQE values are calculated for the following variables:
\begin{itemize}
	\item Wind speed at 700hPa and 850hPa (vector lengths of (u, v) variables at these levels), evaluated at highest quantiles;
	\item Mean sea level pressure departure from standard atmosphere (1013.25 hPa), evaluated at lowest quantiles;
	\item 500hPa geopotential and 850hPa temperature anomaly, evaluated at both lowest and highest quantiles and averaged.
\end{itemize}
From \autoref{fig:rqe}, it is apparent but unsurprising that GraphCast, while excelling in RMSE and ACC metrics, demonstrates the most substantial underestimation of extreme values among all models. This outcome is expected because of the extensive autoregressive training employed by GraphCast. Interestingly, FourCastNet also under-predicts extremes on many variables, possibly indicative of limited modeling capability. Pangu-Weather exhibits a pronounced underestimation of extreme wind speed, while performing better in predicting extremely low-pressure systems. 

The results for Kunyu-Legacy are largely unexpected, showing little underestimation on extreme wind speed, but the worst overestimation on low pressure and anomaly extremes. This suggests the model's tendency to predict extreme events are possibly not impeded by 4-step autoregressive training. However, this approach comes at a cost, as the model produces more unstable outputs for longer lead times.

A noteworthy discovery in these charts is that ECMWF HRES, while mostly neither underestimates nor overestimates wind speed and low pressure extremes, significantly overestimates anomalies at long lead times. This phenomenon also largely explains why Pangu-Weather and Kunyu-Legacy exhibit similar tendencies on these anomalies, since the ERA5 dataset is generated by running an adapted version of ECMWF IFS \cite{hersbach2020}, possibly inheriting similar biases. Theoretically, continuous data assimiliation should enhance dataset self-consistency, but it's unclear to what extent it actually mitigates these subtle model biases embedded in the dataset.

Most significantly, Kunyu, benefiting from training examples at long lead times from the replay buffer, performs more stably across all RQE variables. Despite still underestimating extremes on some variables such as 850hPa wind speed and 500hPa geopotential anomaly, likely due to the presence of regression loss, and over-predicting 700hPa wind speed extremes, Kunyu's RQE curves generally stay closest to zero among all models, including ECMWF HRES.
\subsection{Tropical Cyclones}
Different from previous sections, as there could be potential information leakage from the future in the ERA5 reanalysis, all DLWP models use the same initial condition from ECMWF HRES to ensure comparability. Consequently, GraphCast is substituted by its operational version, refered to as GraphCast-Oper in this paper, capable of taking HRES analysis as input.
\subsubsection{Cyclogenesis and Intensity Forecasting: Case Study of Hurricane Otis}
\begin{figure}[!t]
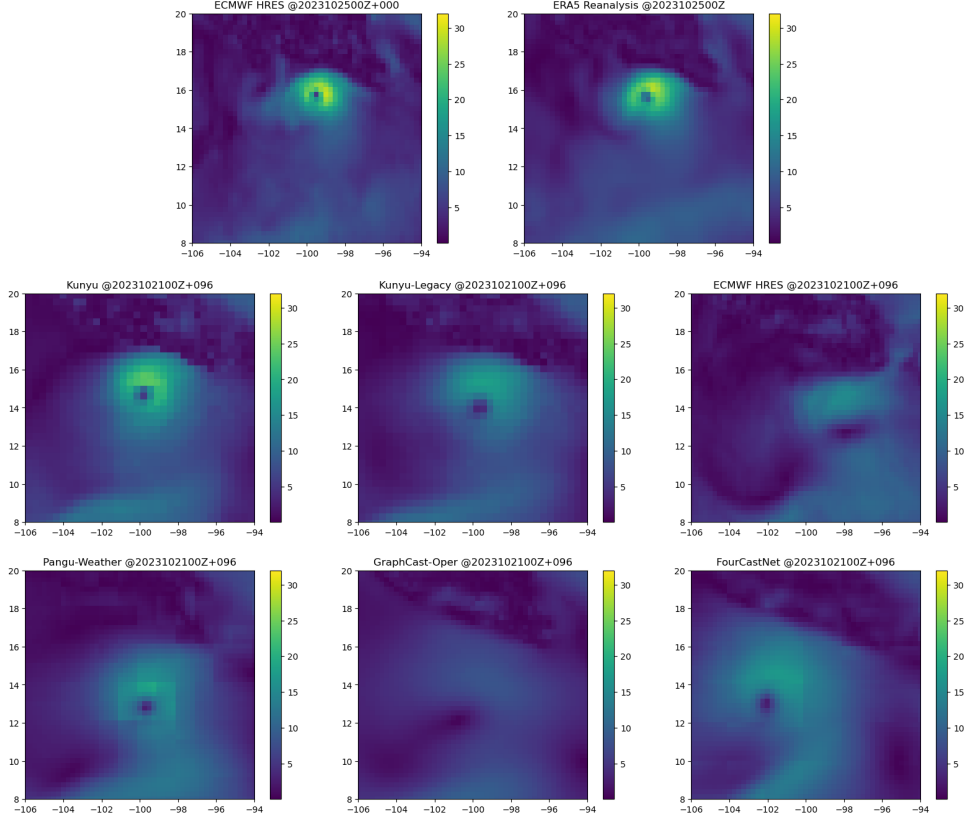

	\centering
	\begin{tabular}{cc}
		\includegraphics[width=4cm]{otis\_hres\_an.png}&\includegraphics[width=4cm]{otis\_era5.png}
	\end{tabular}
	\begin{tabular}{ccc}
		\includegraphics[width=4cm]{otis\_kunyu.png}&\includegraphics[width=4cm]{otis\_kunyu\_legacy.png}&\includegraphics[width=4cm]{otis\_hres.png}\\
		\includegraphics[width=4cm]{otis\_pangu.png}&\includegraphics[width=4cm]{otis\_graphcast.png}&\includegraphics[width=4cm]{otis\_fourcastnet.png}
	\end{tabular}
	\caption{Analyses and forecasts of 850hPa wind speed (unit: m/s) for Hurricane Otis at 00 UTC on October 25. The predictions are provided by Kunyu, Kunyu-Legacy, ECMWF HRES and other DLWP models, initialized with the HRES analysis from four days prior.}
	\label{fig:otis}
\end{figure}
Hurricane Otis was an exceptionally powerful Pacific hurricane which landed near Acapulco on October 25, 2023 at its peak intensity, featuring maximum sustained winds of 165mph/145knots. This hurricane caused severe destruction in the city, earning the distinction of being the strongest landfalling Pacific hurricane in recorded meteorological history. Prior to landfall, Otis underwent explosive intensification, strengthening from a tropical storm to a Category 5 hurricane within a mere 18 hours, according to the National Hurricane Center.

Despite its remarkable intensification and impact, global numerical weather prediction models made notably poor predictions in advance. When Otis formed on October 22, none of the models captured its rapid intensification, with many predicting the storm to dissipate off the Mexican coastline. In \autoref{fig:otis}, DLMP models are executed based on the ECMWF HRES initial condition at 00 UTC on October 21, and the forecasts are then compared with those from ECMWF HRES, as well as the analyses at 00 UTC on October 25, which was when Otis reached Category 5 intensity.

\autoref{fig:otis} clearly reveals the inaccuracy of ECMWF HRES in predicting Hurricane Otis's development. The model predicts merely a shallow and poorly organized low-pressure area over the Eastern Pacific Ocean. GraphCast-Oper fares even worse, presenting only a closed circulation, a result largely anticipated due to the strong impact of autoregressive training. Other regression-based models perform better, forecasting an intensity of a low-end tropical storm.

The standout performer among all models is unequivocally Kunyu, projecting mid-to-high-end tropical storm winds in its forecast. Although still vastly underestimating Otis's actual intensity, it's noteworthy that this underestimation is also evident in the ERA5 reanalysis and even the ECMWF HRES analysis. Reanalysis datasets, owing in part but not solely to low resolution \cite{schenkel2012}, lack a precise estimate of tropical cyclone intensity. Consequently, despite adversarial training, Kunyu, trained on ERA5 reanalysis data, still faces limitations in the quality of intensity estimates. Nevertheless, when compared to other models, including ECMWF HRES, Kunyu's prediction in this case stands out as the closest one. Kunyu's prediction is also the most accurate in pinpointing the storm's location, most likely due to a closer estimate of its intensity, given the correlation between the steering level of a tropical cyclone and its intensity. Other models predicting weaker intensities also position the storm more southerly, whereas the actual Category 5 hurricane was situated more northerly than all model predictions.

The case of Hurricane Otis clearly underscores the inherent limitations of regression in such scenarios. Regression unavoidably causes the underestimation of intensity, especially in situations with high uncertainty, which in turn diverges the storm track. The number of autoregressive steps further exacerbates the shortcomings in both intensity and track. What may seem more surprising is the poor performance of ECMWF HRES, even falling behind most regression-based DLWP models. Actually based on my observations of ECMWF HRES forecasts, such underperformance is not uncommon. The model often lags in predicting the formation and development of tropical cyclones, trailing behind even regression-based models by one or two days. This observation is supported by quantitative evaluations \cite{halperin2013}, where ECMWF HRES exhibits the highest success ratio but the lowest probability of detection among all models except NOGAPS.

In this case, Kunyu, bolstered by adversarial training, appears to correct this model bias and makes much more aggressive, and notably, closer-to-reality predictions on tropical cyclone formation and intensity compared to ECMWF HRES and other regression-based models. This underscores the potential of adversarial training to enhance the predictive capabilities of DLWP models in scenarios characterized by rapid cyclone development.
\subsubsection{Cyclone Track Forecasting: Case Study of Typhoon Khanun}
\begin{figure}[!t]
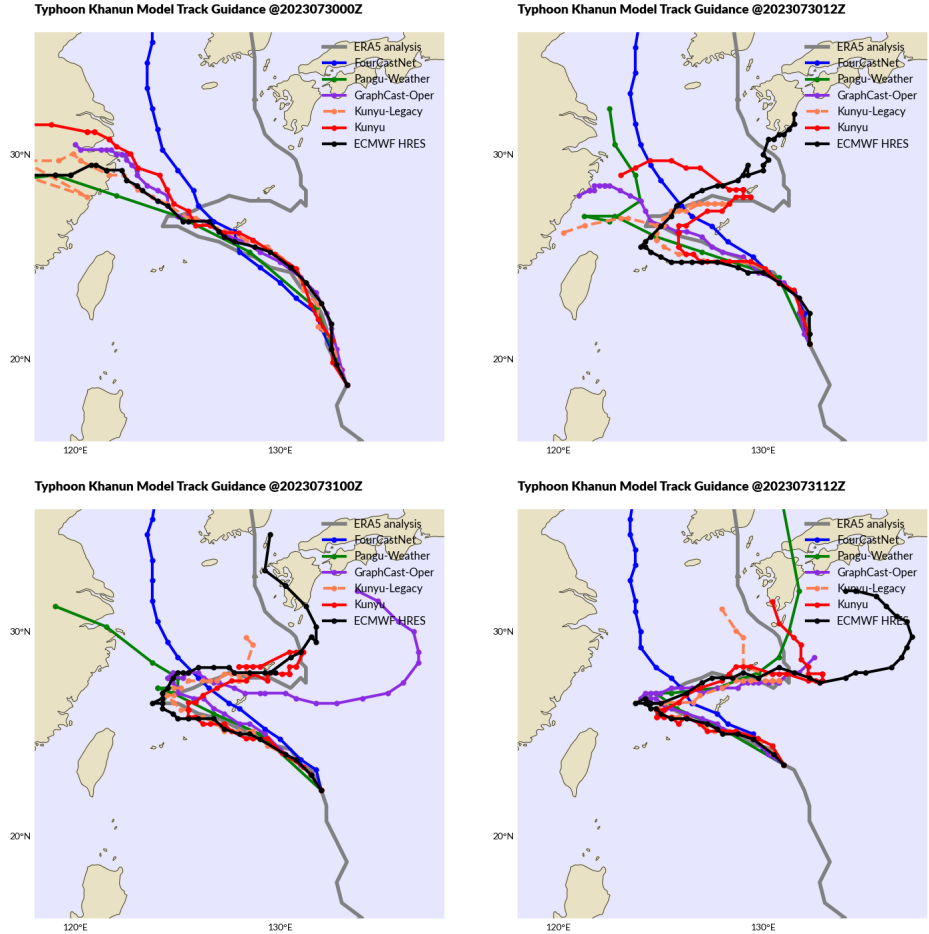

	\centering
	\begin{tabular}{cc}
		\includegraphics[width=6cm]{khanun\_track\_2023073000.png}&\includegraphics[width=6cm]{khanun\_track\_2023073012.png}\\
		\includegraphics[width=6cm]{khanun\_track\_2023073100.png}&\includegraphics[width=6cm]{khanun\_track\_2023073112.png}
	\end{tabular}
	\caption{Track predictions for Typhoon Khanun by Kunyu, Kunyu-Legacy, ECMWF HRES and other DLWP models, with initial conditions derived from HRES analyses at various timestamps. The typhoon's center is determined by the straightforward way of identifying the lowest pressure among surrounding grid points. Tracking stops when the central pressure rises past 998hPa, the estimated pressure when Khanun attained tropical storm intensity.}
	\label{fig:khanun}
\end{figure}
In the 2023 Typhoon season, Typhoon Khanun stands out as a powerful tropical cyclone navigating the most erratic track characterized by two turns over the East China Sea and the Pacific Ocean. This case study investigates how DLWP models and ECMWF HRES respond to changing atmospheric conditions in predicting Typhoon Khanun. The DLWP models are executed on four ECMWF HRES initial conditions from 00 UTC on July 30 to 12 UTC on July 31, and the forecasts are then compared with those from ECMWF HRES and the actual track provided by ERA5 reanalysis.

From \autoref{fig:khanun}, it can be clearly concluded that FourCastNet produces the least accurate track predictions, persisting with a typical parabolic path, likely due to its limited modeling capability. All other models initially predict a landfall in eastern China, but by 12 UTC on July 30, there are some signals indicating a potential deviation. Kunyu, Kunyu-Legacy and ECMWF HRES are the first to switch to a turning scenario near the Ryukyu Islands. In contrast, Pangu-Weather forecasts a turn near the Chinese coastline, and GraphCast-Oper sticks to a landfall, but nearly stalling before reaching the coast. By the next major run at 00 UTC on July 31, the first turn is also shown in the GraphCast-Oper forecast, and the second turn starts to take shape. Kunyu and ECMWF HRES provide the closest estimates of the turning location, while in Kunyu-Legacy's forecast Khanun moves too slowly after the first turn, and GraphCast-Oper predicts the opposite. 12 hours later, Pangu-Weather finally aligns its prediction with Kunyu and Kunyu-Legacy and switches to a two-turn scenario, while ECMWF HRES deviates more from the actual track, and GraphCast-Oper strangely forecasts a dissipation south of Japan.

Generally speaking, the track predictions generated by Kunyu are, at the very least, on par with those from ECMWF HRES, with Kunyu-Legacy closely trailing as the second-best performer. Other regression-based DLWP models respond slower to new signals present in the initial conditions, resulting in a turn occuring one or two major runs later than in ECMWF HRES and Kunyu. The inherent smoothness embedded in the forecasts of regression models likely plays a role, introducing biases in translation speed estimates, especially when faced with high uncertainty and the absence of a strong steering system. An alternative explanation is that regression models tend to smooth out new signals in favor of lower regression losses, making them more inclined to stick to old track predictions. While the smoothness introduced in regression models can be an advantage in scenarios featuring one single strong steering system -- common in most cases where tropical cyclones follow a parabolic path along the side of the steering subtropical ridge -- it proves less effective in cases requiring a more delicate approach to handling subtle atmospheric signals, as demonstrated by the case of Typhoon Khanun.
\section{Discussion}
Kunyu stands out as, to the best of my knowledge, the first DLWP model trained beyond regression loss. Exhibiting impressive performance on extreme weather and tropical cyclone forecasting, while maintaining competitive quantitative forecast skills, Kunyu shows substantial potential for DLWP models to completely replace traditional NWP models. The model's trained ability to forecast a broad array of atmospheric variables has enabled the publication of real-time forecasts on https://kunyu.dev, utilizing initial conditions provided by the ECMWF MARS license, albeit with an approximately 18-hour delay compared to ECMWF HRES.

The impressive performance of Kunyu can't be achieved without its unique model architecture and training process. The combination of spherical convolution and transformer blocks proves highly effective, demanding significantly fewer computational resources for training than competing models like Pangu-Weather and GraphCast. This efficiency enables personal users, like myself, to train a high-performing weather model. Even Kunyu's all-regression counterpart, Kunyu-Legacy, demonstrates promising results with limited regression effects and good forecast skills.

Despite Kunyu has achieved notable successes, certain limitations still persist. One significant issue is the slight lag in forecast skills against ECMWF, especially in terms of ACC. The most likely solution involves switching to larger models and allocating more computational resources, a task that may lie beyond my capabilities. Beyond quantitative metrics, there is also room of improvement to further reduce the effect of regression, which may necessitates a stronger discriminator model or an extensive parameter tuning process. It's important to note that achieving sharper and more detailed outputs might compromise forecast skills, requiring a delicate balance.

In addition to performance-related challenges, Kunyu suffers from artifacts at longer lead times, which should be attributed to the model structure. These artifacts are particularly visible in isobar plots on the Kunyu website, where isobars become more curvy and noisy with increasing lead times, especially in polar areas. Addressing these artifacts is a complex task, with both spherical convolution and transformer blocks potentially contributing to this issue. One possible solution involves manually filtering out higher frequencies, such as keeping spherical harmonics instead of grid point values at each time step -- a method similar to that used in ECMWF IFS \cite{ecmwf2023}. However, this approach would require compatible input and terrain information instead of a simple bilinear interpolation of ERA5 or ECMWF HRES analyses. Alternatively, adopting a reduced gaussian grid similar to ECMWF IFS may suppress artifacts, particularly in polar areas, but such modification would require heavy adjustments on the model architecture, given the requirements of regular grid block layouts for upsampling, downsampling and attention blocks under ordinary implementations.

Besides the model itself, the evaluation results highlight limitations in current training approaches, applicable not only to Kunyu but to the broader DLWP field. The ERA5 dataset, despite its utility, has shortcomings such as limited resolution and significant underestimation of tropical cyclone strengths. Additionally, biases from ECMWF IFS likely persist in the dataset and pose challenges in mitigation, particularly without training on very long time steps. The case study of Typhoon Khanun underscores the influence of input conditions on model performance, emphasizing the need for a data-driven data assimiliation process to push model performance towards new limits.
\section{Acknowledgements}
I would like to give special thanks to the following:
\begin{itemize}
	\item Lambda GPU Cloud, for providing cost-effective and accessible GPU instances for personal users, without a cumbersome and often impossible approval process which is common in other cloud platforms. Without this platform this project would not only incur much higher expenses but also experience significant delays.
	\item Zhang Hao from the Chinese weather enthusiasts community, for creating the https://kunyu.dev website which displays real-time forecast results generated by Kunyu.
\end{itemize}
I would also like to thank the ECMWF Data Support Team for approving the MARS license, my colleagues Ke Liu and Jinyue Su and Shuchang Zhou from MegVii for providing insights on machine learning, Kuo-Feng Chang from the Central Weather Administration of Taiwan and Kian Huang from the University of Utah for providing insights on meteorology and numerical weather prediction.
\printbibliography
\end{document}